\documentclass[fleqn,12pt]{wlscirep}
\title{Knowledge Consensus in complex networks: the role of learning}
\usepackage{multirow}

\author[1,*]{Zhong-Yan Fan}
\author[2]{Ying-Cheng Lai}
\author[1]{Wallace K. S. Tang}
\affil[1]{Department of Electronic Engineering, City University of Hong Kong, Hong Kong}
\affil[2]{School of Electrical, Computer and Energy Engineering, Arizona State University, Tempe, AZ 85287, USA}
\affil[*]{zyfan4-c@my.cityu.edu.hk}
\begin{abstract}

To reach consensus among interacting agents is a problem of 
interest for social, economical, and political systems. A computational
and mathematical framework to investigate consensus dynamics on complex 
networks is naming games. In general, naming is not an independent process 
but relies on perception and categorization. Existing works focus on consensus 
process of vocabulary evolution in a population of agents. However, in order 
to name an object, agents must first be able to distinguish objects according 
to their features. We articulate a likelihood category game model (LCGM) to 
integrate feature learning and the naming process. In the LCGM, self-organized 
agents can define category based on acquired knowledge through learning and use likelihood estimation to distinguish objects. The information communicated 
among the agents is no longer simply in some form of absolute answer, but 
involves one's perception. Extensive simulations with LCGM reveal that a more complex knowledge makes it harder to reach consensus. We also find that agents with larger degree contribute more to the knowledge formation and are more likely to be intelligent. The proposed LCGM and the findings provide new insights into the emergence and evolution of consensus in complex systems in general.
 
\end{abstract}

\begin{document}
\flushbottom
\maketitle
\thispagestyle{empty}

\section{Introduction}

To reach consensus among a population of agents is a problem with significant
applications in social, economical, and political systems. A mathematical and 
computational paradigm to describe, characterize, and understand consensus 
dynamics is naming game (NG) and its variants~\cite{DallAsta:2006(1),DallAsta:2006(2),Barrat:2007Chaos,Lu:2009,lou2018local,Wu2011,LWLCW:2011,Steels:2012,Li:2013CNSNS,Treitman2013,Gao:2014,Mistry:2015,Lou:2015,Niu:2017,Fu:2017,lou2018communicating,zhou2018multi,Li:2017}.
In an NG, agents attempt to reach consensus about a certain object or event via interactions. 
Most existing works emphasize on the consensus of vocabulary among the 
agents~\cite{DallAsta:2006(1),DallAsta:2006(2),Barrat:2007Chaos,Lu:2009,lou2018local,Wu2011,LWLCW:2011,Steels:2012,Li:2013CNSNS,Treitman2013,Gao:2014,Mistry:2015,Lou:2015,Niu:2017,Fu:2017,lou2018communicating,zhou2018multi}.
While vocabulary is often related to objects and plays an important role in 
the development of natural language, for intelligent agents, another 
determining factor in NG is the contents and features of the objects. In fact, 
a pivotal cognitive ability of human being is categorization that recognizes, 
characterizes, and eventually names objects according to their features. 
In our daily activities, majority of the naming actions are 
based on a category rather than a specific term or vocabulary. For example, 
the word ``cat'' refers to a category in which objects have same features 
rather than a specific cat ``Tom'' or ``Kitty.'' For NG to better describe 
consensus dynamics in the real world, it is necessary to incorporate
catagorization through learning and knowledge growth into the model. While 
there were previous works on categorization games that deal with feature 
recognition and analysis~\cite{Steels:2005,Puglisi:2008,Mukherjee:2011,Cui:2014,Baronchelli:2015},
the process of categorization itself remains to be an outstanding topic of 
research, due to the complexity of the underlying process. It is thus a 
challenging task to incorporate perception and categorization into NGs. The 
aim of this paper is to introduce a new kind of category game model to address 
this problem.

In previous works on NGs, consensus is achieved through some learning processes solely governed by interactions between the agents~\cite{Steels:2012}. Therefore, topological features of the agents' interaction network, such as degree, clustering coefficient, and path distance, have significant influences, and it is concluded that higher degrees, lower clustering and shorter network distance tend to promote consensus ~\cite{Barrat:2007Chaos,DallAsta:2006(1),DallAsta:2006(2),Lu:2009,lou2018local}. The knowledge transfer in NGs is not limited between two agents, for example, Li et al. \cite{Li:2013CNSNS} suggested a model with multiple hearers while Gao et al ~\cite{Gao:2014} considered negotiation to take place between multiple indirectly connected agents. An agent can also play the roles of a speaker and a hearer simultaneously~\cite{Gao:2014}. The more general setting was also studied in which the agents possess different propensities such as commitment~\cite{Niu:2017,Mistry:2015} and stubbornness~\cite{Wu2011,Treitman2013}, with the finding that the learning behaviors of these agents can affect significantly the consensus 
dynamics over the whole population. In addition, other issues for more realistic agents, such as learning 
errors during interactions~\cite{Lou:2015}, memory loss~\cite{Fu:2017} and multi-word/language~\cite{lou2018communicating,zhou2018multi}, have been investigated. These works provide great 
insights into consensus dynamics. However, since only vocabulary is focused, significant feature-based cognitive abilities such as categorization were not taken into account which, as natural intuition 
would suggest, may play a more significant role in the emergence and evolution of consensus.  

There were a few proposals on categorization games.
For example, discrimination and guessing games were designed to 
accomplish the task of categorization~\cite{Steels:2005}. In a
discrimination game, a speaker is trained with ground truth so that it 
can relate an object to the actual category using a classifier. Guessing game is 
similar to a typical NG, however, the hearer not 
only acquires the name but also updates his/her classifier for the category 
as instructed by the speaker. Based on these two types of games, the problem 
of color categorization was studied~\cite{Steels:2005}, and a category game 
model (CGM) was developed~\cite{Puglisi:2008}. In a game defined by CGM, 
a pair of objects are presented to a speaker and a hearer, and the target 
topic (the object to be learned) is selected by the speaker. If both objects 
are distinguishable to the speaker, one is randomly chosen as the target 
topic. Otherwise, the speaker must discriminate the two objects by creating 
a new boundary between them before selecting one as the target topic. The
interaction between the speaker and the hearer then makes it feasible to learn
and name the target topic. In the CGM framework, factors such as language 
aging~\cite{Mukherjee:2011}, persistence~\cite{Cui:2014}, and individual 
biases~\cite{Baronchelli:2015} can be studied. A feature common to both 
discrimination game and CGM is that certain pre-requisites are needed. 
Specifically, in a discrimination game, it is necessary to relate the object 
to an actual category, while in CGM, the two presented objects are assumed 
to belong to two different categories. In the real world, it often occurs that
a population can reach an agreement without being given any ground truth, and this has consequences. For example, different language systems can give 
different color categorization, and the consensus of opinions among practitioners in a financial market can trigger a herd behavior that leads to
the fat-tail distributions in prices and 
returns~\cite{Eguiluz:2000}.

To understand and exploit consensus dynamics as realistically as possible
requires the development of a more comprehensive type of game models centered
about features. The base of our model is the recently proposed domain learning 
naming game (DLNG)~\cite{Li:2017} to solve the color categorization problem 
through elimination of pre-requisites in the learning process. The color 
perception process in DLNG follows a variant algorithm based on the majority 
rule. Agents in DLNG have numerous sensors uniformly distributed in the domain, and one sensor can be dominated by at most one category. Then, an object is deemed as in a category if the majority of sensors near the object belong to that category. However, the original DLNG is a kind of coarse-grained model where 
a determined category may contain distinct objects. For example, intelligent 
agents such as humans are generally capable of assessing that crimson is 
much redder than magenta, although both colors belong to the same category 
of red. 

To overcome this difficulty, here we propose a likelihood category 
game model (LCGM), in which self-organized agents define category based 
on acquired knowledge and use likelihood estimation to distinguish the 
objects in the same category. The information communicated among the agents 
is no longer simply some absolute answer but involves agents' perception. 
That is, the agents are able to classify objects in terms of distinct 
likelihoods as determined by their knowledge and update knowledge through 
learning new objects. For the special case where the learning domain is
highly localized, our model reduces to a variant of the minimum NG. The primary contribution of this paper is the proposed LCGM, with a new concept of likelihood to allow agents assess an object via perception, which better reflect our real experiences. Moreover, conclusive remarks related to consensus characteristics are provided. Firstly, it is noticed that short distance and heterogeneity can facilitate the consensus. Secondly, LCGM clearly reflects a more complex knowledge makes it harder to reach consensus. Lastly, agents with larger degree contributes more to the knowledge formation and consensus and accordingly to be more “intelligent”. Our work in LCGM provides novel insights into the process of consensus, with potential applications to predicting and controlling consensus -- problems with implications to social, economical, and even political systems where achieving consensus is often a desired objective.

\section{Model}

\subsection{Agent model} 
An object $o_i$ in our LCGM is represented by a point in an $N$-dimensional 
domain~\cite{Puglisi:2008}, that is
$x_i \equiv \left(x_{i,1},x_{i,2},\cdots,x_{i,N}\right) \in D \subseteq \Re^N$. Each agent $j$ maintains a set of categories $\mathcal{C}_j$ in its memory. 
Each category, $c_k \in \mathcal{C}_j$, is described by a weight $\omega_k$ and $N$'s normal 
distributions $\mathcal{N}(\mu_{k,n},\sigma_{k,n}^2)$  for 
the corresponding $n$ dimension with $n=1,2,\cdots,N$. An agent, 
$\mbox{Agent}_j$, classifies the object $o_i$ based on a likelihood score  
that characterizes how likely $o_i$ belongs to category $c_k$. The score can 
be calculated through
\begin{equation} \label{eq1}
LS_j(o_i,c_k) = g(\omega_{k}) \prod_{n=1}^{N} f\left(\frac{x_{i,n} 
- \mu_{k,n}}{\sigma_{k,n}}\right) / f(0)
\end{equation}
where $f(\cdot)$ is the probability density function (PDF) of the standardized 
normal distribution and $g(\cdot)$ is a scale function defined as
\begin{equation} \label{eq2}
g(\omega) = \tanh(\alpha \omega),
\end{equation}
where $\alpha$ is a constant. To be concrete, we set $\alpha = 2.5$, so  
$g(1)=0.987 \approx 1.0$. Equation~(\ref{eq1}) indicates that the likelihood 
score contains two components: prior and likelihood. The prior component, 
represented by $g(\omega_{k})$, indicates the probability that category $c_k$ 
is selected for an arbitrary object. The likelihood component is specified by 
the product term in Eq.~(\ref{eq1}). The normalized value 
$f\left(\frac{x_{i,n} - \mu_{k,n}}{\sigma_{k,n}} \right) / f(0)$ 
measures the likelihood of $o_i$ belonging to category $c_k$ along the $n$th 
dimension. In the absence of any correlation among different dimensions, the 
product term gives the joint probability value. 

In our model, each agent can execute four possible activities: object 
identification, category updating, category creation, and category deletion,
which are described, as follows. 

\begin{figure}[ht]
\centering
\includegraphics[width=\linewidth]{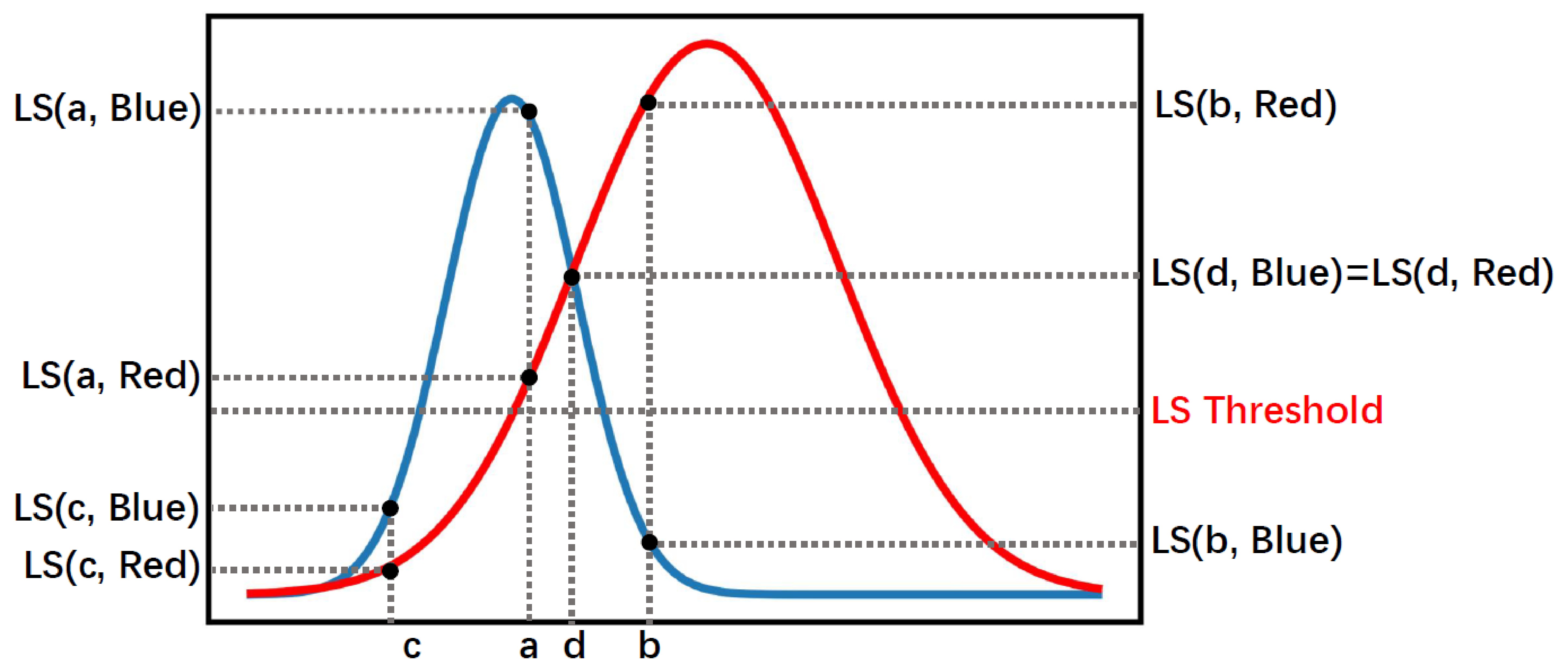}
\caption{ {\bf An illustrative example of object identification in 
one dimension}. Suppose the agent has two categories: ``Blue'' and ``Red''.  
With regard to objects ``a'', ``b'', ``c'' and ``d'', there are four 
cases ($a$, $b$, $c$ and $d$, respectively). In case $a$, since 
$LS(a,\mbox{Blue}) > LS(a,\mbox{Red})$ and $LS(a,\mbox{Blue}) > LST$, the 
agent returns the name of category ``Blue" and $LS(a,\mbox{Blue})$. In 
case $b$, $LS(b,\mbox{Red}) > LS(b,\mbox{Blue})$ and $LS(b,\mbox{Red})> LST$, 
the name of category ``Red'' and $LS(b,\mbox{Red})$ are returned. In case 
$c$, $LS(c,\mbox{Blue})$ is higher but still lower than LST, so the agent 
returns $\mbox{name}=\mbox{null}$ and $LS=0$. In case $d$, 
$LS(d,\mbox{Red}) = LS(d,\mbox{Blue}) > LST$, resulting in category ``Red'' 
or ``Blue'' being randomly picked. Case $d$ is rare because it occurs only 
at the intersection point of the two curves.}
\label{fig:2}
\end{figure}

\underline{\it Objective identification activity (OIA).} 
Given an object $o_i$, $\mbox{Agent}_j$ returns $LS_j(o_i,c_a)$ and the 
corresponding $\mbox{name}_a$, where 
$c_a =\arg \underset{c_x \in \mathcal{C}_j} \max \ LS_j(o_i,c_x)$ is the 
category having the highest score. For $\mathcal{C}_j = \emptyset$, 
$\mbox{Agent}_j$ will return a null name ($\mbox{name}_a = \mbox{null}$) 
and a zero likelihood score [$LS_j(o_i,c_a) = 0$]. In addition, if $LS_j(o_i,c_a)$ is below a 
specified threshold, denoted as $LST$, then the agent will also return a 
null name and a zero likelihood score. An illustrative 
example of OIA for the one-dimensional case is given in Fig.~\ref{fig:2}.

\begin{figure}[ht]
\centering
\includegraphics[width=\linewidth]{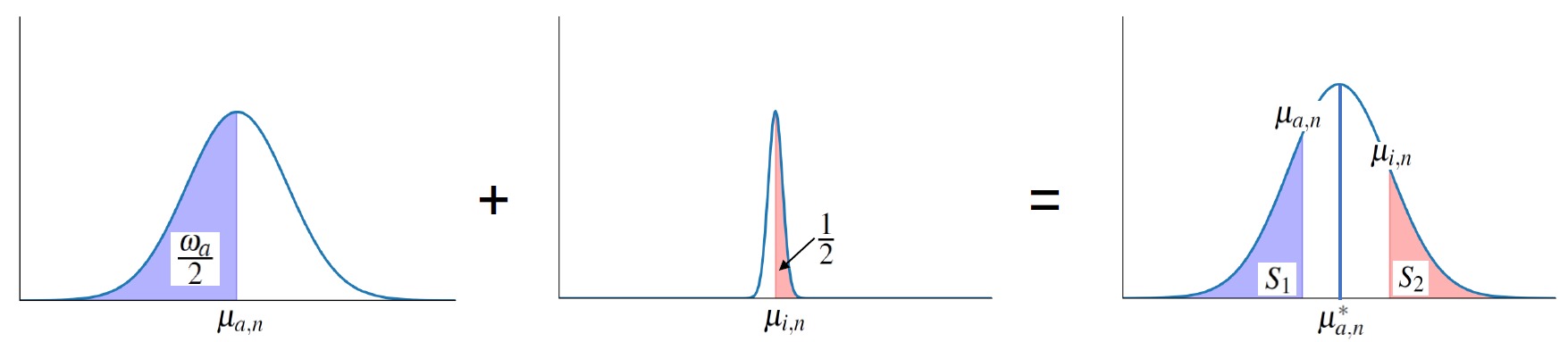}
\caption{ {\bf An illustrative example of $\sigma$ updating}. For each 
$n$th dimension, the category $c_a$ (left) and object $x_i$ (middle) are 
merged into a new category $c_a^{*}$ (right). Let the parameters be $\omega_a=7$ 
and $\sigma_a=1$ for $c_a$, and $\omega_i=1$ for the object. 
Equation~(\ref{eq3}) thus gives $\omega^*_a=\omega_a + \omega_i = 8$.
For the $n$th dimension, the shaded areas of $c_a$ and $x_i$ are 
$\omega_a/2 = 3.5$ and $\omega_i/2 = 0.5$, respectively. 
Consider $S_1 = \omega^*_a \times Pr(x_{j,n} \leq \mu_{a,n})$ and 
$S_2 = \omega^*_a \times Pr(x_{j,n} \geq \mu_{i,n})$ in the updated 
category, where $Pr(\cdot)$ stands for the probability. There are two cases 
with Eq.~(\ref{eq10}). First, if $\mu_{a,n} = 0$ and $\mu_{i,n} = 0.8$, 
then Eq.~(\ref{eq4}) gives $\mu_{a,n}^{*} = 0.1$ and $\sigma_{a,n}^{*}$ is 
kept at the same value as $\sigma_{a,n}$ 
[$\sigma_{a,n}^{*} = \max(0.456,0.636,1) = 1$]. Correspondingly, one has 
$S_1 = 3.68 > \omega_{a}/2$ and $S_2 = 1.94 > 1/2$. Second, 
if $\mu_{a,n} = 0$ and $\mu_{i,n} = 2$, then one has $\mu_{a,n}^{*} = 0.25$ 
and $\sigma_{a,n}^{*} = \max(1.141,1.589,1) = 1.589$. As a consequence, one
has $S_1 = 3.5 = \omega_{a}/2$ and $S_2 = 1.083 > 1/2$. 
Equation~(\ref{eq10}) stipulates that the inequalities 
$S_1 \geq \omega_a/2$ and $S_2 \geq 1/2$ hold. Note 
that object $x_i$ (middle) should be a Dirac delta function 
($\sigma_{i,n} \to 0$) at $\mu_{i,n}$, and it is plotted as a bell-shape 
only for visual clarity.}
\label{fig:13}
\end{figure}

\underline{\it Category updating activity (CUA).} 
If $o_i$ is assigned to category $c_a$, the features 
($\mu_{a,1}$, $\sigma_{a,1}, \cdots, \mu_{a,N}$, $\sigma_{a,N}$ and 
$\omega_{a}$) of $c_a$  will be updated by merging the information of $o_i$. 
Since $o_i$ is solely represented by the point $x_i$ in the domain $D$, it
is reasonable to let the weight of $x_i$ be one. As a result, $\omega_a$ 
is increased by one, forming a new weight
\begin{equation} \label{eq3}
\omega_{a}^{*} = \omega_{a} + 1
\end{equation}
For each $n$th dimension, the updated mean, $\mu^{*}_{a,n}$, is the 
weighted average of the category $c_a$ and the point $x_i$, which can be 
obtained through
\begin{equation} \label{eq4}
\mu^{*}_{a,n} = \frac{\omega_{a} \times \mu_{a,n} + x_{i,n}}{\omega_{a}^*}
\end{equation}
Finally, the updated $\sigma^{*}_{a}$ can be obtained from the following 
equation:
\begin{equation} \label{eq10}
\sigma_{a,n}^{*} = \max\left(\frac{-|x_{i,n} - \mu_{a,n}^{*}|}{F^{-1}\left(\frac{1}{2\omega_a^*}\right)}, \frac{-|\mu_{a,n} - \mu_{a,n}^{*}|}{F^{-1}\left(\frac{\omega_{a}}{2\omega_a^*}\right)}, \sigma_{a,n}\right),
\end{equation}
where $F(.)$ is the CDF of $f(\cdot)$. An illustrative example is given in 
Fig.~\ref{fig:13}. Note that $\sigma$ of the category distribution is 
non-decreasing because only exceptions will result in its changes while 
normal observations will have no effect on $\sigma$.

\underline{\it Category creation activity (CCA).} 
If an object $o_i$ with $\mbox{name}_i$ concluded in a game cannot be related 
to any existing category associated with the agent, a new category $c_a$ is 
created. The parameters $\sigma_a$ and $\omega_a$ are set at some default 
values, while $\mu_a = x_i$ and $\mbox{name}_a = \mbox{name}_i$.

\underline{\it Category deletion activity (CDA).} 
In reality, if knowledge is not recalled, it will be forgotten gradually. 
Similarly, in LCGM, category will be removed from one's memory if there is no updating, i.e.,
it has not been learned for a long time. The removal of a category is based 
on its weight. For any time $t$, the weight of category $c_k$, $\omega_k$ is updated as 
\begin{equation} \label{eq8}
\omega_{k,t} = \omega_{k,t-1} \times e^{-\phi}
\end{equation}
where $\phi$ is the ``forgetting'' factor that describes the ``forgetting 
speed'' of a category in the agent's memory, $\omega_{k,t}$ and $\omega_{k,t-1}$ are the value of $\omega_{k}$ at time $t$ and $t-1$, respectively.

As Eq.~(\ref{eq1}) indicates, a reduction in $\omega_k$ will also affect 
the chance of category $c_k$ being assigned in OIA. If $\omega_{k,t}$ 
further reduces and becomes smaller than a pre-defined threshold (denoted 
as $\omega_{Th}$), the category $c_k$ will be removed from the memory of 
the agent. However, once an object falls into category $c_k$, $\omega_k$ 
becomes larger than one [Eq.~(\ref{eq3})] via CUA.

\subsection{Game rules.} 
Game participants are all agents in a population on a network. Initially, 
every agent has empty memory: $\mathcal{C}_j =\emptyset,\ \forall j$. A 
pair of connected agents, $\mbox{Agent}_A$ and $\mbox{Agent}_B$, are randomly selected
in a game that proceeds as follows.

(1) An object $o_i$ is presented to both agents. Via OIA, $\mbox{Agent}_A$ 
returns $\mbox{name}_a$ and $LS(x_i,c_a)$, while $\mbox{Agent}_B$ returns 
$\mbox{name}_b$ and $LS(x_i,c_b)$. 

(2) $\mbox{Agent}_A$ and $\mbox{Agent}_B$ conclude the name of $o_i$, 
referred to as $\mbox{name}_i$, based on the following rules:
if $LS(o_i,c_a) > LS(o_i,c_b)$, $\mbox{name}_i = \mbox{name}_a$ and 
$\mbox{Agent}_A$ wins the game; 
if $LS(o_i,c_a) < LS(o_i,c_b)$, $\mbox{name}_i = \mbox{name}_b$ and 
$\mbox{Agent}_B$ wins the game; 
if $LS(o_i,c_a) = LS(o_i,c_b) = 0$, $\mbox{name}_i$ is randomly generated 
and the game is a draw; 
if $LS(o_i,c_a) = LS(o_i,c_b) \neq 0$, a winner is randomly selected. 
For example, if $\mbox{Agent}_A$ is the winner, one has 
$\mbox{name}_i = \mbox{name}_a$.
In addition, if $\mbox{name}_a = \mbox{name}_b \neq \mbox{null}$, the game is 
successful and $\mbox{name}_i = \mbox{name}_a$. Otherwise, it has failed. 

(3) $\mbox{Agent}_A$ and $\mbox{Agent}_B$ update/create their categories 
according to a set of rules. The rules for $\mbox{Agent}_A$
can be described as follows (similar for $\mbox{Agent}_B$).
If $\mbox{name}_i$ equals the name of a category, say 
$c_k \in \mathcal{C}_A$, CUA is operated on category $c_k$. However,
if $\mbox{name}_i$ is not contained in any category in $\mathcal{C}_A$, 
CCA is carried out to create a new category with $\mbox{name}_i$ and 
$\mu_i = x_i$. 

(4) All agents in the population delete expired categories via CDA. 

\begin{figure}[ht]
\centering
\includegraphics[width=0.8\linewidth]{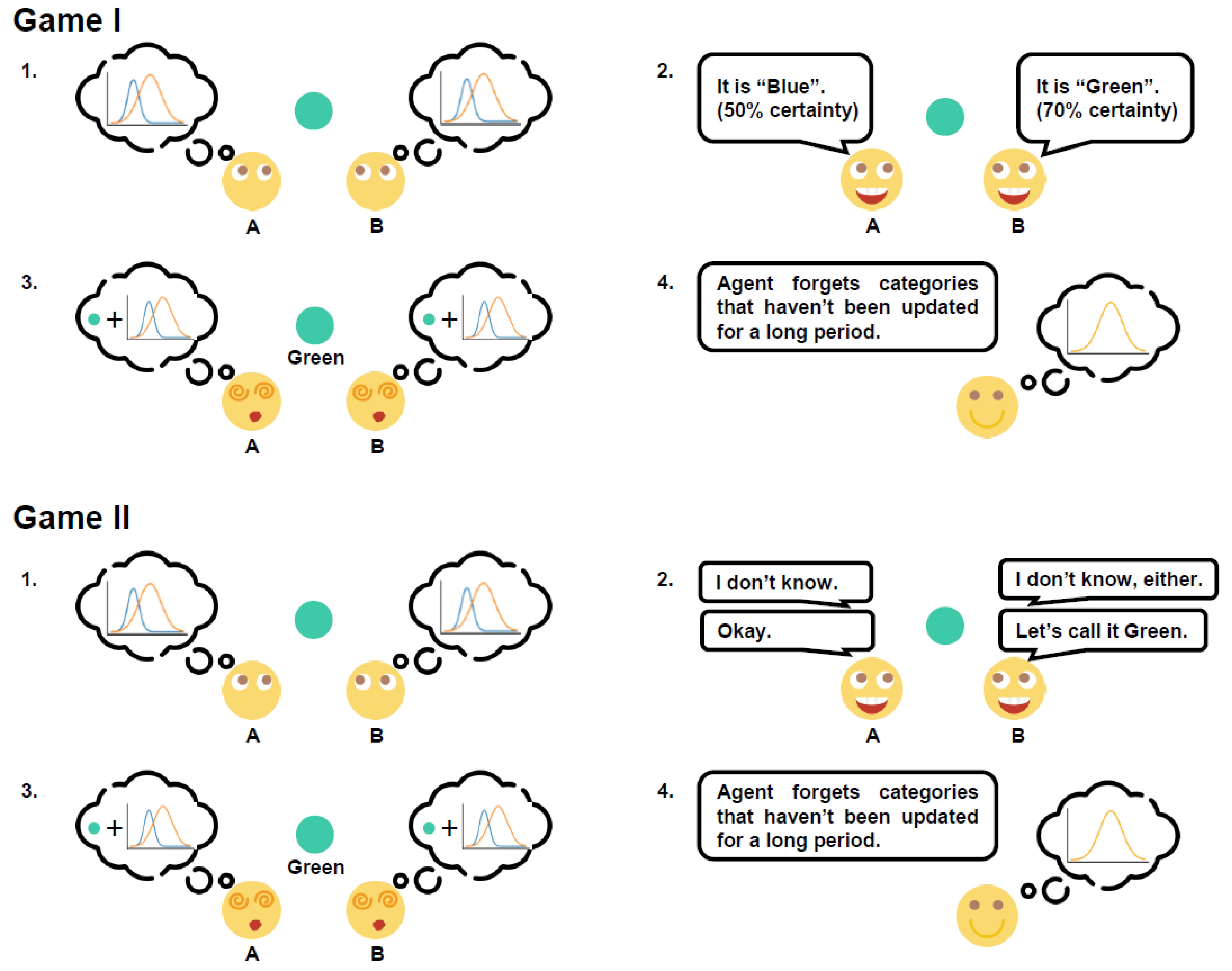}
\caption{ {\bf Two illustrative examples of game process in terms of 
color objects}. In Game I, via OIA, $\mbox{Agent}_A$ considers that the 
object is ``Blue'' with 50\% certainty, while $\mbox{Agent}_B$ regards
it as ``Green'' with 70\% certainty. As $\mbox{Agent}_B$ is more confident, 
the object is concluded as ``Green.'' The two agents then update their 
memories based on the new knowledge that the object is named ``Green.'' 
In addition, both agents forget categories that have not been recalled 
for a long period of time. In Game II, neither agent can identify the 
object. The agents return $\mbox{name}=\mbox{null}$ and $LS=0$ via OIA. 
As a result, a new name, say ``Green,'' is assigned. The remaining steps 
[Steps (3) and (4)] are the same as those in Game I.}
\label{fig:1}
\end{figure}

Two illustrative examples of the game process are given in Fig.~\ref{fig:1}.

\section{Methods}

We focus on color categorization to demonstrate and analyze LCGM in the
present study. Color objects are randomly generated in $RGB$ form with
sample size $256 \times 256 \times 256$, and are then mapped into $CIELab$ 
color space for agents. Specifically, the color objects and categories are 
represented in the $CIELab$ color space which is three dimensional: 
(L*, a*, b*), where L* stands for the lightness, a* and b* are the two 
abstract opponent dimensions~\cite{Hunter:1958}. The $CIELab$ color space 
is assumed to be homogeneous, i.e., the difference between two color objects 
is defined as the Euclidean distance between them, as specified by the year 
1976 version of $CIELab$ color space. (Remark: To better fit with human  vision, $\Delta E$ was re-defined in the year 2000 version~\cite{Luo:2001} where the color space is no longer homogeneous. However, this does not affect our main results.)

LCGM is applied to a population of agents connected through a pre-defined
network, either from the real world or from a specific model. Initially, each 
agent has an empty memory. Games are conducted as 
described in {\it Game Rules}, and repeated until a predefined number 
of iterations is reached. For all the simulations in this paper, the iteration
number is $10^7$.

Model parameters are set as follows. 
\begin{enumerate}
\item \vspace*{-0.1in} 
Scaling factor in Eq.~(\ref{eq2}): $\alpha = 2.5$
\item \vspace*{-0.1in}
Default value of $\sigma$ in CCA: $\sigma_{default} = 5.0$. 
Empirically, two colors with the value of $\Delta E$ between 1 and 10
appear similar in human vision, so we set $\sigma_{default} = 5.0$.
\item \vspace*{-0.1in}
Default value of $\omega$ in CCA: $\omega_{default} = 1.0$. 
\item \vspace*{-0.1in}
The threshold parameter $\omega_{Th}$: a category will be deleted during 
CDA if its weight $\omega$ is smaller than $\omega_{Th}$. We set 
$\omega_{Th}=0.01$.
\item \vspace*{-0.1in}
Likelihood score threshold (LST): If the LS computed from an object-category 
pair is smaller than LST, the object should not belong to that category. 
We choose different values for LST to investigate its impact on the learning process.
\item \vspace*{-0.1in}
Forgetting factor $\phi$: it characterizes the speed at which a category 
is forgotten. We choose different values of $\phi$ to investigate its impact on the learning process. 
\end{enumerate}

We conduct simulations by applying LCGM to two social networks (a subgraph of Facebook network and a subgraph of E-mail network) and various heterogeneous networks. The results from the Facebook network and the E-mail network are averaged by 10 runs. To investigate the effect of network topology, artificial networks are constructed. For each setting, ten network realizations are used to calculate the various statistical averages.

\section{Results}

To demonstrate LCGM, we test it on learning color categorization by a 
population of agents. We define the following performance metrics.

\textbf{Accuracy (ACC)} indicates the success rate of the games, which is 
computed for every $K$ games and defined as:
\begin{equation} \label{eq5}
ACC = \frac{k_{success}}{K}
\end{equation}
where $k_{success}$ is the number of successful games in every $K$ iterations. 

\textbf{Number of categories (NC)} specifies the number of categories 
maintained by each agent, which reflects the resolving power of agents after 
learning. The average number of categories ($NC_{Avg}$) is primarily concerned in this paper. 

\textbf{Total number of distinct names (TDN)} records how many distinct names 
remained in the whole population.

\textbf{Consensus score (CS)} characterizes the consensus of the whole 
population. A set of sampled objects ($\mathcal{O}$) is firstly selected. In this work, $\mathcal{O}$ consists of 512 objects uniformly sampled from the domain $D$. These objects are  
presented to every pair of agents in turns (including agents who are not neighbors) and games are performed. For $\mbox{Agent}_i$ and 
$\mbox{Agent}_j$ with $i \neq j$, $CS_{i,j}$ is defined as 
\begin{equation} \label{eq6}
CS_{i,j} = \frac{agm_{i,j}}{|\mathcal{O}|}
\end{equation}
where $agm_{i,j}$ is the total number of successful games that $\mbox{Agent}_i$ 
and $\mbox{Agent}_j$ made for all $o_i \in \mathcal{O}$, and $|\cdot|$ is the 
cardinality of a set. The $CS$ of the entire population $\mathcal{V}$ is the average over all possible agent pairs, which can be computed as
\begin{equation} \label{eq7}
CS = \frac{\sum_i{\sum_{j\neq i}{CS_{i,j}}}}{|\mathcal{V}| \times (|\mathcal{V}| - 1)}
\end{equation}

\subsection{Application: consensus in social network}

We apply our LCGM algorithm to two social networks, a subgraph of Facebook social network~\cite{SNAP} and a subgraph of an E-mail network~\cite{SNAP_Email}, collected by the SNAP laboratory of Stanford university. The Facebook subgraph has 1,034 nodes and 26,749 edges. Its clustering coefficient, average path length and degree-based Gini coefficient are $0.5$, $2.9$ and $0.48$, respectively. The E-mail network has 986 nodes and 25,552 edges. Its clustering coefficient, average path length and degree-based Gini coefficient are $0.31$, $2.6$ and $0.56$, respectively. The relatively large value of the clustering coefficient and short average path length are typical of small-world networks, while the relatively high values of the Gini coefficient suggest that the networks are inhomogeneous.

Typical simulations with $LST = 0.1$ and $\phi = 0.00002$ were conducted. On Facebook social network, ACC reaches 88.0\%, the average NC and TDN are about 4.7 and 6.8, respectively. And on E-mail network, ACC reaches 88.6\%, the average NC and TDN are 4.6 and 5.8, respectively.

However, the high accuracy does not guarantee full consistencies of agents under the framework of category game. Since accuracy only considers gaming actions occur between two mutual neighbors, consensus between agents who are not neighbors within the population is ignored. The consensus score (CS) is then used to characterize the consistency of the whole population. The average CS for Facebook social network and E-mail network are about 0.71 and 0.85, respectively.

\begin{figure}[ht]
\centering
\includegraphics[width=160mm]{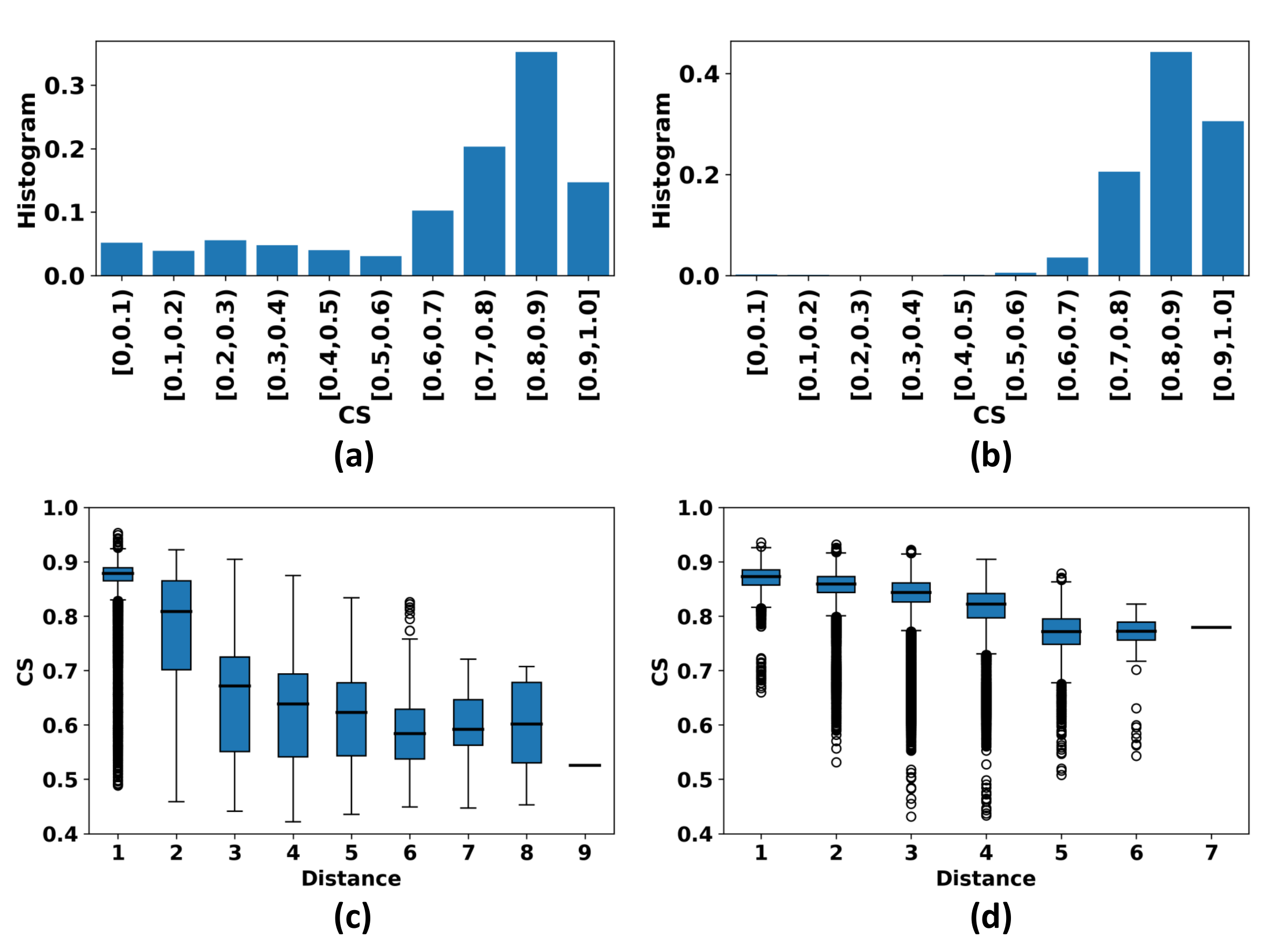}\\
\caption{ {\bf Consensus score (CS) of agent pairs}. Shown are (a) the histogram of the consensus score of Facebook social network, (b) the histogram of the consensus score of E-mail network, (c) boxplot of the correlation between distance and CS of Facebook social network and (d) boxplot of the correlation between distance and CS of E-mail network.}
\label{fig:4}
\end{figure}

The performance of CS was further investigated. The histograms of consensus score are plotted in Fig.~\ref{fig:4}~(a)~and~(b). The majority of agent pairs have CS higher than 0.8 for both networks. However, some agent pairs have relatively low CS, particular for the Facebook social network. Then, we analyzed the influence of distance of agent pairs and observed a negative correlation between CS and distance of agent pairs (See Fig.~\ref{fig:4}~(c)~and~(d)). It is because agents involved in the transmission path would incorporate their understandings to the information. As a result, a longer distance between two agents generally imposes a higher variation in the knowledge between them.

The adoption of LCGM on the  two social networks revealed that agents basically reach the consensus through the mutual learning, with consensus score 0.71 (the Facebook network) and 0.85 (the E-mail network) respectively. While the CS of the E-mail network is much higher than that of the Facebook network even though they have similar ACC. The extended analysis on CS and distance implies that the shorter average path length of the E-mail networks may facilitate the consensus. Other topology properties, such as clustering and heterogeneity, will be discussed later in this section.

Based on LCGM, an interesting question is, among the agents, who more frequently takes the lead in the 
game by providing the answer? The leader is the winner of the game as 
specified by the game rules. To identify the leader, we define the following
game score for each agent: 
\begin{equation} \label{eq5a}
GS(A) = \sum_{i}GS_{i}(A)
\end{equation}
where $GS_i(A)$ is the game score obtained by $\mbox{Agent}_A$ in $i$th game:
\begin{equation} \label{eq6a}
GS_i(A) =  
\begin{cases}
      1, & \text{if $\mbox{Agent}_A$ participated in the $i$th game and won} \\
      0.5, & \text{if $\mbox{Agent}_A$ participated in the $i$th game and the result was a draw}\\
      0,& \text{otherwise.}
\end{cases}
\end{equation}
Equation~(\ref{eq6a}) indicates that our proposed learning game is 
effectively a positive-sum game. In each game, the increment of the total 
score is one. When an agent wins the game, it dominates the game solely and 
takes the whole score. When the game is a draw, both players score 0.5 as 
they contribute to knowledge formation equally. In this case, the loser 
receives no penalty for the reason that learning should not be discouraged. 
To eliminate the effect of opportunity earning (agents involved in more games 
likely will have higher scores), we further define the following game scoring rate:
\begin{equation} \label{eq7a}
GSR(A) = \frac{GS(A)}{\#\mbox{Games}(A)}
\end{equation}
where $\#Games(A)$ is the total number of games involving $\mbox{Agent}_A$. 

\begin{figure}[ht]
\centering
\includegraphics[width=160mm]{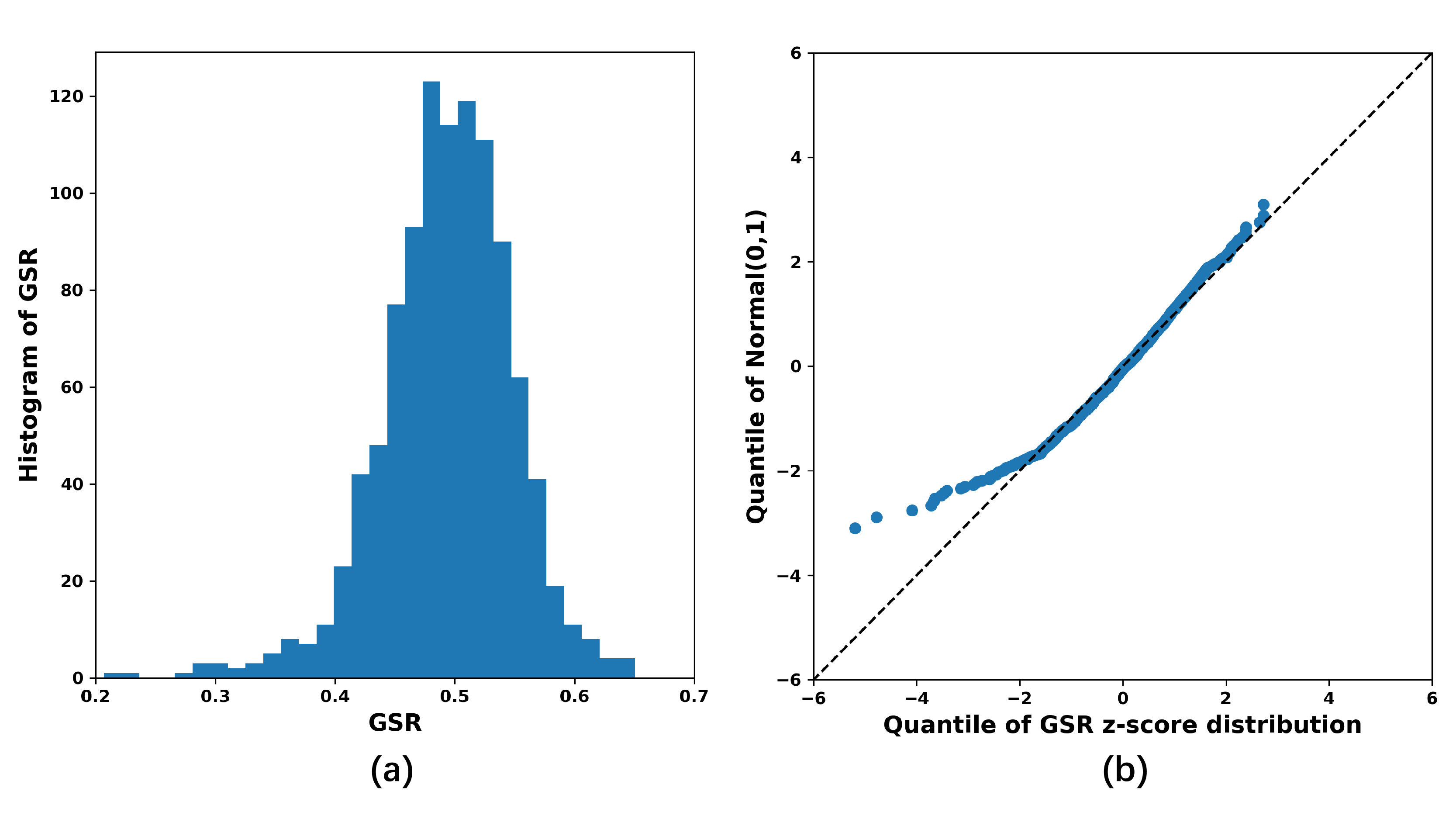}\\
\caption{ {\bf Statistical features of game scoring rate for the 
Facebook social network}. Shown are (a) the histogram of the game 
scoring rate (GSR) and (b) the corresponding 
Quantile-Quantile plot (QQ plot) for $\mu_{GSR} = 0.49$ and 
$\sigma_{GSR}$ = 0.06.}
\label{fig:5}
\end{figure}

As it is similar for both studied social networks, we only discuss the results of the Facebook social network here. Figure~\ref{fig:5}~(a) shows the GSR distribution, giving a bell-shape in which the majority (93.5\%) of agents have GSR values 
between 0.4 and 0.6. To check whether the distribution is normal,  
we use the Quantile-Quantile plot (QQ plot). Specifically, a dot $(x,y)$ in 
the QQ plot means $F_{X}(x) = F_{Y}(y)$, where $F_{X}(x)$ is the cumulative 
density function of random variable $X$. If two distributions are identical, 
all dots would be located along the diagonal. 
As shown in Fig.~\ref{fig:5}~(b), 
dots within $x \in (-2,2)$ are quite close to the diagonal, indicating 
that the GSR possesses the characteristic of a normal distribution in 
this range. Such a bell-shape distribution reflects most agents 
have moderate intelligence after sufficient learning and communications. Furthermore, the emergence of a light right tail and a heavy left tail observed in Fig.~\ref{fig:5}~(b) indicates that ``geniuses are minority'' and 
``dummies are more than you expected.''

\begin{figure}[ht]
\centering
\includegraphics[width=160mm]{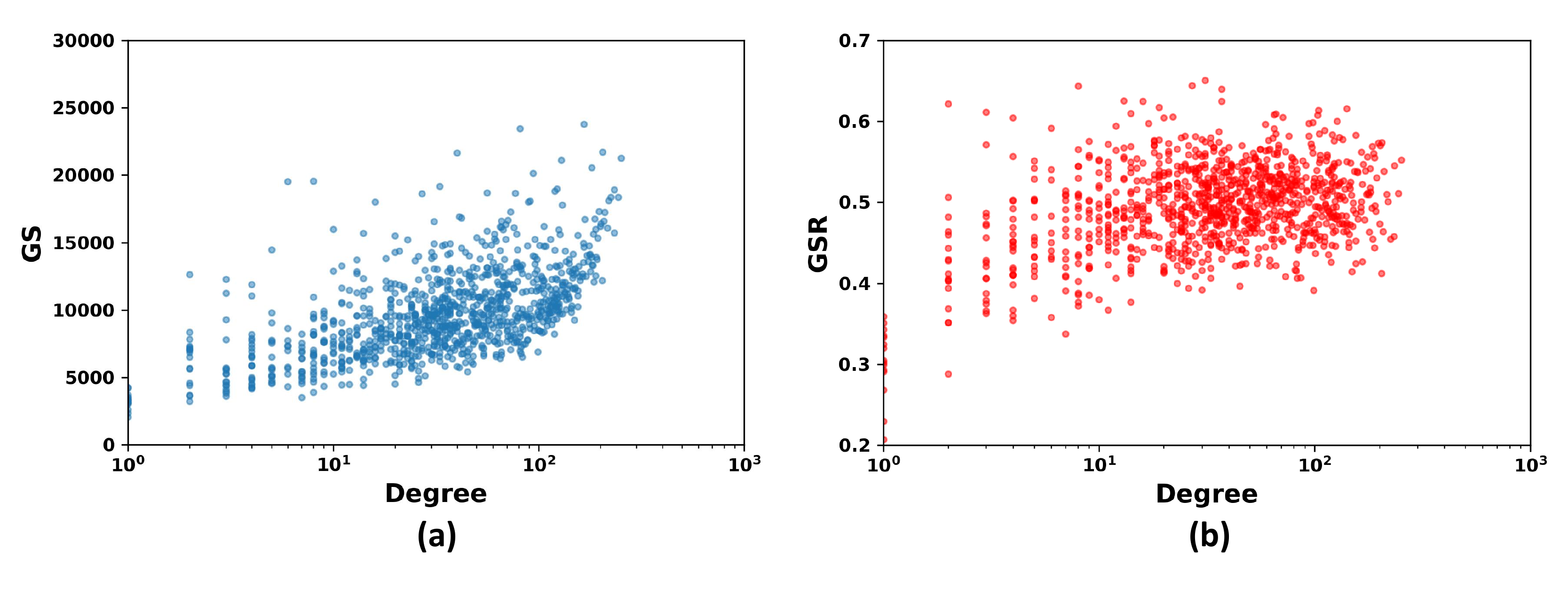}\\
\caption{ {\bf Correlation between GS/GSR and degree centrality for the 
Facebook social network}. The Pearson 
correlation coefficients are $\rho_{log(d),GS}=0.60$ and 
$\rho_{log(d),GSR}=0.40$.}
\label{fig:6}
\end{figure}

The correlation between GS/GSR and degree is shown in 
Fig.~\ref{fig:6}. As defined in~Eqs.~(\ref{eq5a}) and (\ref{eq6a}), GS 
reflects the frequency of transmitting self-knowledge to others. 
Figure~\ref{fig:6}(a) shows that agents with large degrees are likely to have 
a high GS, implying that agents with larger degree contributed more to the consensus process. Different from GS, GSR measures the capability of scoring 
in a game [Eq.~(\ref{eq7a})]. Figure~\ref{fig:6}(b) reveals a positive 
correlation between GSR and degree, indicating that agents with larger 
degrees are more ``intelligent'' with respect to eventual knowledge 
formation through more active learning and communication.

\subsection{Effects of Likelihood score threshold (LST) and forgetting factors (FF)}

Then, the effects of likelihood score threshold (LST) and forgetting factors (FF) of categories were investigated. Again, the studies were based on the Facebook social network.

{\it Likelihood score threshold of agents}. As shown in Fig.~\ref{fig:2}, 
LST imposes the minimum requirement for classifying an object into a 
category. Accordingly, a high LST value encourages the generation of a new 
category. To investigate the effect of LST, we simulate the learning process for different values of LST with fixed forgetting factor $\phi = 0.00002$ in Eq.~(\ref{eq8}).

\begin{figure}[ht]
\centering
\includegraphics[width=160mm]{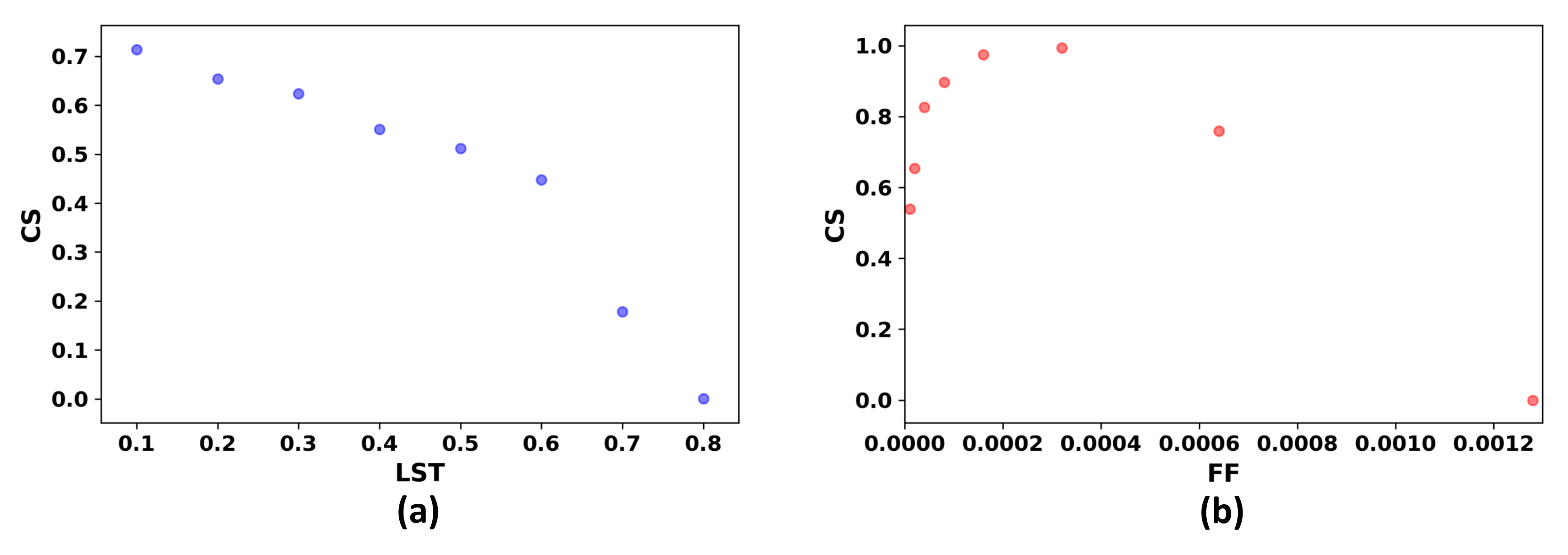}\\
\caption{ {\bf CS of simulations with different LST/FF}.}
\label{fig:7}
\end{figure}

As shown in Table~\ref{tab:1}~ (left), agents with high LST (the ``rigorous'' agents) spontaneously possess more categories. And the more complex knowledge makes it harder to reach consensus, resulting a lower accuracy. This is further confirmed by the plot of consensus scores with different LST as shown in Fig.~\ref{fig:7}~(a), which clearly shows that the increment of LST would depreciate CS significantly.

\begin{table}[ht]
\centering
\caption{\label{tab:1} \textbf{Summary of LGCM results (after $10^7$ iterations) for different values of LST/FF} Each data point is the result of averaging over ten game realizations.}
\begin{tabular}{|p{1.5cm}|p{1.5cm}|p{1.5cm}|p{1.5cm}||p{1.5cm}|p{1.5cm}|p{1.5cm}|p{1.5cm}|}
\hline
\multicolumn{4}{|c||}{LST ($\phi = 0.00002$)}&\multicolumn{4}{c|}{FF ($LST = 0.2$)}\\
\hline
LST & ACC & Avg. NC & TDN & $\phi$ & ACC & Avg. NC & TDN \\
\hline
0.1 & 88.0\% & 4.7 & 6.8 & 0.00001 & 81.3\% & 10.2 & 15.4\\
0.2 & 85.5\% & 5.5 & 7.7 & 0.00002 & 85.5\% & 5.5 & 6.7\\
0.3 & 84.1\% & 7.5 & 10.0 & 0.00004 & 89.3\% & 3.6 & 4.3\\
0.4 & 82.3\% & 9.3 & 42.3 & 0.00008 & 93.8\% & 2.2 & 2.2\\
0.5 & 79.4\% & 13.6 & 412.9 & 0.00016 & 98.0\% & 1.3 & 1.3\\
0.6 & 71.9\% & 27.6 & 6472.6 & 0.00032 & 99.9\% & 1.0 & 1.0\\
0.7 & 38.6\% & 154.6 & 67670.0 & 0.00064 & 79.8\% & 3.55 & 1399.6\\
0.8 & 4.2\% & 394.3 & 174366.5 & 0.00128 & 0\% & 6.92 & 3549.2\\
\hline
\end{tabular}
\end{table}

{\it Forgetting factor of categories.}
We next study the mechanism responsible for categories to be gradually
forgotten associated with category deletion activities. In general, the 
threshold for removal of a category, $\omega_{Th}$, should be small because,
if $\omega_{Th}$ is large, even useful categories would be 
deleted, possibly resulting in a dramatic effect on the learning process. However, if the value of $\omega_{Th}$ is too small, the category with small $\omega$ results small likelihood
score, leading to a small probability of any update (learning). 
These empirical considerations lead to our choice of $\omega_{Th} = 0.01$.

To be concrete, we fix $\mbox{LST}=0.2$ and focus on the impact of varying the forgetting factor $\phi$. The results are shown in Table~\ref{tab:1}~(right) and Fig.~\ref{fig:7}~(b). When $\phi$ increases, it becomes hard for agents to remember information, eventually leading to a reduction in the number of categories. Apparently, when there are fewer categories, it is easier to reach consensus, which is verified through the ACC and CS results. However, the correlation between the FF and CS is non-monotonous. For $\phi > 0.00032$, the CS decreases dramatically and even equals to 0 for $\phi=0.00128$ through the process. That is, quite naturally, consensus can never be reached if agents forget things too fast. 

It is remarked that, the average result of CS is given. For the case with $\phi=0.00064$, some experiments reached consensus with only one category (similar to the case with $\phi=0.00032$) while others had $CS=0$ (similar to the case with $\phi=0.00128$).

\subsection{Effect of network topology.}
Does the network topology have a significant effect on LCGM? To address this question, we considered four topological characteristics, namely the average degree, network size, heterogeneity, and clustering.

Based on original BA scale-free networks~\cite{barabasi1999emergence}, the impacts of average degree (with fixed network size) and network size (with fixed average degree) onto the performance of LCGM were investigated. The simulation results show no observable influence on LCGM by varying the average degree (see Supplementary Table S1). While the network size affects the consensus of LCGM slightly (see Supplementary Table S2), network with smaller size can reach a better consensus, which meets most people's intuition that consensus are easily to be reached among fewer agents. 

To investigate the effects of heterogeneity and clustering, two types of heterogeneous networks were adopted: scale-free network with aging~\cite{Zhu:2003} and scale-free network with clustering 
attachment~\cite{Holme:2002}. The first type of networks incorporates the 
aging effect of nodal attraction, and the connection probability is redefined as  
\begin{equation} \label{eq9}
\mbox{Prob}(i) = \frac{\mbox{degree}_{i} \times e^{-\lambda \times \mbox{age}_{i}}}{\sum_{j} 
\mbox{degree}_{j} \times e^{-\lambda \times \mbox{age}_{j}}}
\end{equation}
where $\lambda$ is a tunable parameter reflecting the aging speed of nodal 
attraction and $\mbox{age}_{i}$ denotes the age of node $i$. When node $i$ is newly added, $\mbox{age}_i = 0$. After every nodal addition cycle, $\mbox{age}_i$ is incremented by one. Networks with distinct degrees of homogeneities can be generated by adjusting the value of $\lambda$. 

We perform LCGM on the aging scale-free networks of size 1000 with average degree of 10 
for different values of $\lambda$. As $\lambda$ is increased, 
the Gini coefficient is significantly reduced and the average path length 
increases, but the clustering coefficient is hardly affected. We find that
an increase in the value of $\lambda$ has little effect on the accuracy
($\mbox{ACC}$) and the average number of categories ($\mbox{NC}_{avg}$) 
in LCGM. However, the consensus score (CS) decreases significantly with $\lambda$ even though 
the accuracy remains high (see Supplementary Table S3). As discussed before, there is a negative correlation between the agent-pair distance and CS. Therefore, We further investigated average consensus scores of agent pairs with respect to distance and $\lambda$ (see Supplementary Table S4). It is observed that $\lambda$ has no observable influence on CS between agents with short distance ($\leq2$), while for agent pairs with longer distance, CS decreases generally as the network becomes more homogeneous ($\lambda$ increases).

Then, the effect of clustering on LCGM is investigated by considering the conventional clustering coefficient. Scale-free networks with different clustering coefficients are obtained through the process 
of clustering attachment~\cite{Holme:2002} where an edge is added to connect a new node and 
one of its two-hop neighbors, forming a triangle. The probabilities of 
clustering and preferential attachment are $p$ and $1-p$, respectively.  
If no two-hop neighbor is available for a new node, preferential attachment 
is adopted. We perform LCGM on networks of size 1000 (with average degree of 10) for 
different values of $p$ (See results in Supplementary Table S5). It can be easily observed that the clustering coefficient increases with the value of $p$. An increase in the value of $p$, however, 
has only limited effect on the Gini Coefficient and the average path length. From the results, it can be concluded that the clustering coefficient does not hurt the consensus. Since categories have weights in LCGM, several strong categories will survive through evolution. Different from traditional naming game models, LCGM allows these categories to coexist, and consequently, they will dominate the whole population together.

\section{Discussion}

To reach consensus among a population of diverse agents is a problem
of great complexity but one with broad relevance. A computational and
mathematical paradigm to investigate this problem is naming games. 
Most existing works in this area were based on some learning processes that 
solely rely on direct interactions among the agents without taking 
into account agents' perception. The basic idea underlying our proposed
likelihood category game model (LCGM) is that knowledge acquiring is 
essential to achieving consensus. In LCGM, self-organized agents rely on 
acquired knowledge to define category and employ statistical likelihood 
estimation to distinguish and ``name'' objects that belong to the same 
category. Particularly, the agents equipped with knowledge acquiring 
are capable of exploiting distinct likelihoods as determined by their 
knowledge to classify objects. Importantly, knowledge is not static but 
dynamic: agents update knowledge through learning. The agents in our LCGM
are thus ``smart,'' more closely mimic those in the real social, economical, 
and political world.   

The distinct features of our proposed LCGM are the following. Firstly, it is 
a truly autonomous category game model, eliminating the need for ground truth 
knowledge. Secondly, introducing the concept of likelihood in this context
makes the game model more realistic because, in each game, agents not only 
return the names but also the likelihood scores that represent their 
confidence corresponding to the answer. The information communicated among 
the agents is ``smart'' in the sense that it is no longer restricted 
to some form of absolute answer but is more content or feature based. 
Thirdly, in a pairwise game, the more knowledgeable agent (with a higher 
likelihood score) becomes the ``teacher,'' a feature that fits with the 
interaction in the real world. Through the incorporation of learning, our model 
generalizes the existing naming game models and is closer to describing 
reality. (Our model reduces to a variant of the much studied minimum
name game in the special case where the learning domain is singular in the
sense that its extent is effectively zero, i.e., no learning.) 

Our model provides novel insights into consensus dynamics. For example,
there is a trade-off between the amount of knowledge and consensus, providing
a quantitative explanation for the phenomenon that consensus is hard to be 
reached among serious agents with a high LST (who 
know more categories). Another finding is that hubs contributed more to knowledge formation, which accordingly have a larger probability to become ``smart'' and to take lead in a game. By 
studying the effects of network structural characteristics on the consensus
dynamics, we identify the impacts of distance and heterogeneity on consensus. While the findings are preliminary, they are useful for understanding the dynamical evolution of consensus and may even serve as the base for formulating control strategies to harness consensus dynamics, warranting further efforts in investigating learning and likelihood based games.

\bibliography{main}

\section*{Acknowledgements}

YCL would like to acknowledge support from the Vannevar Bush Faculty
Fellowship program sponsored by the Basic Research Office of the Assistant
Secretary of Defense for Research and Engineering and funded by the Office
of Naval Research through Grant No.~N00014-16-1-2828. ZYF and WKST would like to acknowledge support from City University of Hong Kong through Research Grant No.~7004835.

\section*{Author contributions}
W.K.S.T. and Z.Y.F. conceived research. Z.Y.F. performed computations. 
W.K.S.T. and Z.Y.F. analyzed the results. W.K.S.T., Z.Y.F., and Y.C.L. 
wrote the paper.

\end{document}